\documentclass[seceq]{ptptex}
\usepackage{graphicx} 
\usepackage{latexsym}
\notypesetlogo
\begin{document}
\begin{flushright}
RUP-18-22 \\
 August, 2018
\end{flushright}
\vspace{10mm}
\begin{center}
\Large{ \bf Quark mass function at finite temperature \\ in real-time formalism }
\vspace{15mm}

\large{ Hidekazu {\sc Tanaka} and Shuji {\sc Sasagawa} \\
Department of Physics, Rikkyo University, Tokyo 171-8501, Japan\\
}
 \end{center}

\begin{center}

\vspace{25mm}

{\Large ABSTRACT}
 \end{center}
        
\vspace{10mm}
\def\proj{{\bf P}}  
\def\slsh#1{{#1}{\kern-6pt}/{\kern1pt}}  



        

We investigate properties of quark mass functions at finite temperature in quantum chromodynamics calculated by Schwinger-Dyson equation in real-time formalism without the instantaneous exchange approximation, in which one-loop integration is performed in Minkowski space. In our model, an imaginary part of the mass function is directory evaluated. 
 
  \newpage

\section{Introduction}

Chiral phase transition in quantum chromodynamics (QCD) is one of the most interesting phenomena to be understood.

@So far, the chiral phase transitions have been studied by various methods. One such method is implementation of Schwinger-Dyson equation (SDE)[1,2], which can evaluate nonperturbative phenomena. Many works for chiral symmetry breaking have been done with the SDE in momentum representation, in which one-loop contribution is integrated over Euclidean space.  

Some calculations of the mass function for fermion with the SDE have been done in Minkowski space.  In Ref.[3], a spectral representation for Green functions is  assumed, in which the mass functions are calculated in Lorentz invariant form.  In Ref.[4], explicit one loop contribution of the mass function has been calculated. However the mass function is evaluated only one iteration from a constant initial mass as an input. 

In order to extend the SDE at finite temperature, we need to integrate the energy and the momentum separately due to existence of the Boltzmann factor.

At finite temperature and density for equilibrium systems, the imaginary-time formalism (ITF) is implemented, which continues to Euclidean space at zero temperature limit.

On the other hand, the real-time formalism (RTF) for nonequilibrium systems is formulated in Minkowski space.  The SDE in RTF has been studied with instantaneous exchange approximation (IEA)[5,6], in which gauge boson energy is neglected. In the IEA, the mass function does not depend on the energy. Furthermore, the critical coupling of chiral symmetry breaking in quantum electrodynamics(QED) is about a half of that calculated with four momentum integration in Euclidean space at zero temperature.[6-9] An alternative method has been proposed in Ref. [10], in which an energy independence condition for the mass function is required. In evaluation of the mass function, only results for static limit as momentum $p\rightarrow 0$ with zero energy $p_0=0$ have been presented. [11]
 
 Analytic continuation from Euclidean space to Minkowski space is valid in perturbative calculation if pole positions in complex plane of energy are known. However it is not trivial in nonperturbative region, particularly when the mass function depends on the energy.

So far, the structure of the quark mass function at strong coupling region in entire range of energy and momentum space has not been fully studied in Minkowski space  at finite temperature.

In the previous papers[12,13], we formulated the SDE for QED and for QCD in which the momentum integration is performed in Minkowski space without the IEA.

In this paper, we extend  our previous method to calculate the quark mass function with the SDE in the RTF at finite temperature without the IEA.

In section 2, we formulate the SDE in the RTF without the IEA. In section 3, some numerical results are shown.
 Section 4 is devoted to summary and some comments. 

\section{SDE for quark mass function}

 In the RTF, two types of fields specified by 1 and 2 are implemented in the theory, in which the type-1 field is the usual field and the type-2 field corresponding to a ghost field in the heat bath.

We calculate the 1-1 component of a self-energy of quark $\Sigma^{11}(p)$ in QCD , which is given by
\begin{eqnarray}
 -i\Sigma^{11}(P)=(ig_s)^2C_{\rm F}\int{d^4Q \over (2\pi)^4}\gamma^{\mu}iS^{11}(Q)\Gamma^{\nu}iD^{11}_{\mu\nu}(K),
\end{eqnarray}
in one-loop order, where $S^{11}$ and $D_{\mu\nu}^{11}$ are the 1-1 components of the thermal propagators for a quark with momentum $Q=(q_0,{\bf q})$ and a gluon with momentum $K=P-Q=(k_0,{\bf k})$, respectively. Here, $P=(p_0,{\bf p})$ is an   external momentum of the quark. 
The strong coupling constant and the color factor are denoted by $g_s$ and $C_{\rm F}=4/3$, respectively.

The 1-1 component of the quark propagator in the RTF is given as
\begin{eqnarray}
iS^{11}(Q)  = i\left[(S_{\rm F}(Q))_{\rm R}+i(S_{\rm F}(Q))_{\rm I}\epsilon(q_0)\tanh{q_0 \over 2T}\right] 
\end{eqnarray}
 with a temperature $T$, where we define $\epsilon(z)=\theta(z)-\theta(-z)$ with the step function $\theta(z)$.
Here, $(S_{\rm F}(Q))_{\rm R}$ and $(S_{\rm F}(Q))_{\rm I}$ are the real and the imaginary parts of the quark propagator $S_{\rm F}(Q)$. 
  \footnote{The real and the imaginary terms of a propagator $G$ are defined by 
$$ (G)_{\rm R}={1 \over 2}(G+G^*)$$
and 
$$ i(G)_{\rm I}={1 \over 2}(G-G^*)$$
, respectively.}

In order to explain our algorithm, we consider a simple definition of the quark propagator as
\begin{eqnarray}
iS_{\rm F}(Q)  \equiv i(\slsh{Q}+M(Q)) I_{\rm F}(Q).
\end{eqnarray}
with
\begin{eqnarray}
I_{\rm F}(Q) = {1 \over q_0^2-q^2-M^2(Q)+i\varepsilon}.
\end{eqnarray}

The 1-1 component of the gluon propagator\footnote{We neglect an effective gluon mass and do not separate the gluon propagator to a longitudinal and a transverse parts for simplicity.} 
    in RTF is given as 
 \begin{eqnarray}
iD_{\mu\nu}^{11}(K)=i\left[(D_{{\rm F}\mu\nu}(K))_{\rm R}+i(D_{{\rm F}\mu\nu}(K))_{\rm I}\epsilon(k_0)\coth{k_0 \over 2T}\right], 
\end{eqnarray}
where
 \begin{eqnarray}
iD_{{\rm F}\mu\nu}(K)\equiv\left(-g_{\mu\nu}+{K_{\mu}K_{\nu} \over K^2}\right)iD_{\rm F}(K)
\end{eqnarray}
with
 \begin{eqnarray}
iD_{\rm F}(K)={i \over K^2+ i\varepsilon}. 
\end{eqnarray}
The quark-gluon vertex is defined by $\Gamma_{\mu}=\gamma_{\mu}$. 
Our model corresponds to a mass function in the Landau gauge at $T=0$. 

Integrating over the azimuthal angle of the momentum ${\bf q}$, the trace of the self-energy $\Sigma^{11}$ is given by 
\begin{eqnarray}
M^{11}(P) \equiv{1 \over 4}Tr[\Sigma^{11}(P)] = -{3iC_{\rm F} \over 2\pi^2}\int^{\Lambda_0}_{-\Lambda_0}dq_0 \int^{\Lambda}_{\delta} dq {q \over p} \alpha_s [MI_{11}J_{11}](P,Q) 
\end{eqnarray}
with $p=|{\bf p}|$,$q=|{\bf q}|$ and $\alpha_s=g_s^2/(4\pi)$, where,
\begin{eqnarray}
MI_{11}= (MI_{\rm F})_R+i(MI_{\rm F})_{\rm I}N_{\rm F}(T,q_0),
\end{eqnarray}
and
\begin{eqnarray}
J_{11}=(J_{\rm F})_{\rm R}+i(J_{\rm F})_{\rm I}N_{\rm B}(T,k_0),
\end{eqnarray}
with
$N_{\rm F}(T,q_0)=\epsilon(q_0)\tanh[q_0/(2T)]$ and $N_{\rm B}(T,k_0)=\epsilon(k_0)\coth [k_0/(2T)]$, and
\begin{eqnarray}
J_{\rm F}=\int^{\eta_+}_{\eta_-}dkkD_{\rm F}=\int^{\eta_+}_{\eta_-}dk{k \over K^2+i\varepsilon},
\end{eqnarray}
with $\eta_{\pm}=|p\pm q|$ and $k=|{\bf k}|$, respectively.

The real part $M_R $ and  the imaginary part $M_{\rm I} $ of the mass $M$ are given by 
\begin{eqnarray}
M_{\rm R}=(M^{11})_{\rm R}
\end{eqnarray}
and
\begin{eqnarray}
M_{\rm I}=(M^{11})_{\rm I}/N_{\rm F}(T,q_0),
\end{eqnarray}
respectively.

In Minkowski space, the  propagator $I_F$ in Eq.(2$\cdot$4) rapidly varies near $Q^2=q_0^2-q^2 \simeq (M^2)_{\rm R}$, if the imaginary part of the mass function $(M^2)_{\rm I}$ is small, which is one of the difficulties for numerical calculation in Minkowski space. Therefore, it is necessary to perform integration efficiently.
As implemented in the previous works [7][8], we divide the $q_0$ integration into small ranges and integrate the quark propagator over $q^{(l)}_0\leq q_0 \leq q^{(l+1)}_0~(q_0^{(1)}=-\Lambda_0,q_0^{(N)}=\Lambda_0)$ as
\begin{eqnarray}
M^{11} (P) \simeq -{3iC_{\rm F} \over 2\pi^2} \int^{\Lambda}_{\delta} dq {q \over p}\sum_{l=1}^{N-1}\langle\alpha_s J_{11}(q)\rangle_l [MI_{11}(q_0^{(l+1)},q_0^{(l)})].
\end{eqnarray} 
with
$$
[MI_{11}(q_0^{(l+1)},q_0^{(l)})] = [(MI_{\rm F})(q_0^{(l+1)},q_0^{(l)})]_R 
$$
\begin{eqnarray}
+ i[(MI_{\rm F})(q_0^{(l+1)},q_0^{(l)})]_I N_{\rm F}(T,\langle q_0 \rangle_l) 
\end{eqnarray}
where the remaining contributions of the integrand are averaged for the range $q^{(l)}_0\leq q_0 \leq q^{(l+1)}_0$. 
 Here, $\langle X \rangle_l$ denotes the average of a function $X(q_0)$ at $q_0=q_0^{(l+1)}$ and $q_0=q_0^{(l)}$ as $ \langle X \rangle _l=[X(q_0^{(l+1)})+X(q_0^{(l)})]/2$.
Convergence of the calculations are significantly improved by our method.
The explicit expressions are summarized in Appendix A, in which we implement the running coupling constant $\alpha_s$.

\section{Numerical results}

In this section, some numerical results are presented. We solve the SDE presented in Eq. (2$\cdot$14) by a recursion method starting from a constant mass at $T=0$. \footnote{The initial input parameters are $M_{\rm R}=\Lambda_{\rm QCD}$ and $M_{\rm I}=0$ at $T=0$ with $\Lambda_0=\Lambda=10\Lambda_{\rm QCD}$ and $\delta=0.1\Lambda_{\rm QCD}$ with $\varepsilon=10^{-6}$. In evaluation of  the quark mass function at $T+\Delta T$, we implement the solution of $M$ obtained at $T $  as the initial input.
 Here, we define $M= M_{\rm R}+i M_{\rm I}$. }

First, we evaluate an effective quark mass for resonance peak of the effective quark propagator at $T=0$.

The denominator of the effective quark propagator is written as
\begin{eqnarray}
D_F(P)= (P^2-(M^2)_R)^2+((M^2)_I)^2 
\end{eqnarray}
in our approximation, which gives the resonance peak at $ P^2=(M^2)_{\rm R}$. 

For each iteration, we calculate the quark mass function normalized as 
\begin{eqnarray}
M^{(n+1)}(P^2) = m(\zeta^2)+M^{(n)}(P^2)-M^{(n)}(\zeta^2), 
\end{eqnarray}
where $n$ denotes the number of iterations. Here, the mass function is normalized by a current quark mass at large $\zeta^2$, in which perturbative calculations are  reliable. In the iteration, the mass function $M(p_0,p)$ in integrand of   Eq.(2$\cdot$14) is replaced by the renormalized one obtained by the previous iteration. Here, $m(\zeta^2)$ is a renormalized mass at a renormalization scale $\zeta$.\footnote{We take $m(\zeta^2)=0$ at $\zeta=10\Lambda_{\rm QCD}$.  }

Here, $M(P^2)$ is defined as 
\begin{eqnarray}
M(P^2) = M(p_0,\delta)\theta(|p_0|-\delta) 
\end{eqnarray}
for $P^2>0$ with   $p=\delta$, where $P^2=p_0^2-p^2$.

In Fig.1, the dependences on $P^2$ for the mass function $(M^2)_{\rm R}$ with $\Lambda_{\rm QCD}=0.30{\rm GeV},$ $0.32{\rm GeV}$ and  $0.35{\rm GeV}$ are presented at $T=0$. A dashed line denotes $ (M^2)_{\rm R} = P^2$. Therefore, the crossed point corresponds to a position of the resonance peak of the effective quark propagator. 

As shown in Fig.1, the peak position of the fermion propagator depends on the QCD parameter $\Lambda_{\rm QCD}$, which are roughly  $\sqrt{(M^2)_{\rm R}}\simeq \Lambda_{\rm QCD}$ at $ P^2=(M^2)_{\rm R}$ in our model.

In order to search for critical point in which the chiral symmetry is restored, we evaluate 
\begin{eqnarray}
\langle |M|\rangle =\int^{\Lambda_0}_{-\Lambda_0}dp_0\int^{\Lambda}_{\delta}dp|M (p_0,p)|,
\end{eqnarray}
\begin{eqnarray}
\langle M_{\rm R}\rangle =\int^{\Lambda_0}_{-\Lambda_0}dp_0\int^{\Lambda}_{\delta}dp(M (p_0,p) )_{\rm R}
\end{eqnarray}
and 
\begin{eqnarray}
\langle M_{\rm I}\rangle =\int^{\Lambda_0}_{-\Lambda_0}dp_0\int^{\Lambda}_{\delta}dp(M (p_0,p) )_{\rm I}
\end{eqnarray}
as order parameters. 

\begin{figure}
\centerline{\includegraphics[width=10cm]{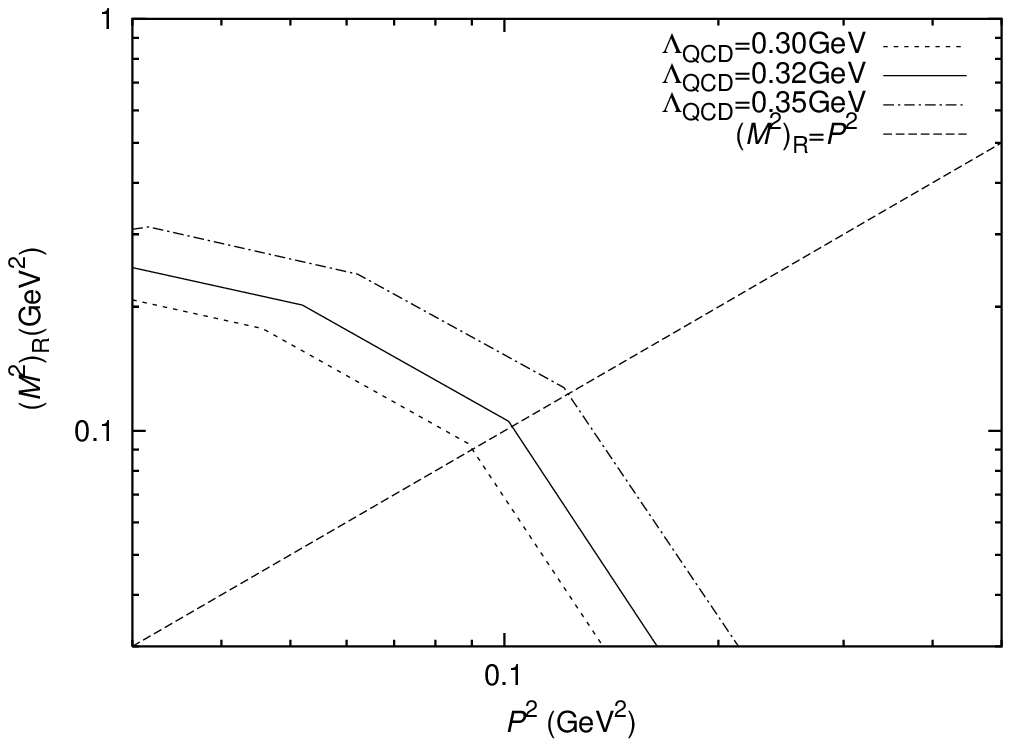}}
\caption{ The $P^2$ dependences of $(M^2)_{\rm R}$ for $P^2>0$ with  $p_0>\delta$ at $T=0$. The dashed line denotes $ (M^2)_{\rm R} = P^2$. }
\end{figure}

In Fig.2, the $T$ dependences of $\langle |M|\rangle $ for $\Lambda_{\rm QCD}=0.30{\rm GeV}, 0.32{\rm GeV}$ and $0.35{\rm GeV}$ are presented.
The critical temperature of the transition $T_{\rm C}$ depends on the QCD scale parameter $\Lambda_{\rm QCD}$. In our model, $0.3 {\rm GeV} \leq \Lambda_{\rm QCD}\leq 0.35{\rm GeV}$ gives $0.16 {\rm GeV} \leq T_{\rm C} \leq 0.19 {\rm GeV}$.

In Fig.3, the $T$ dependences of the integrated mass functions $\langle |M|\rangle, \langle M_R\rangle, \langle M_I\rangle $ are presented at $\Lambda_{\rm QCD}=0.32 {\rm GeV}$, which gives $T_{\rm C}\simeq 0.17{\rm GeV}$ with $\sqrt{(M^2)_{\rm R}}\simeq 0.32{\rm GeV}$ for $P^2=(M^2)_{\rm R}$ at $T=0$.

As shown in Fig.3, the imaginary part of the mass function $\langle M_{\rm I}\rangle $ is non-zero value for broken chiral symmetric phase below  the critical temperature  $T_{\rm C}$, which means the massive quark state may be unstable if energy scale  rapidly changes.

\begin{figure}
\centerline{\includegraphics[width=10cm]{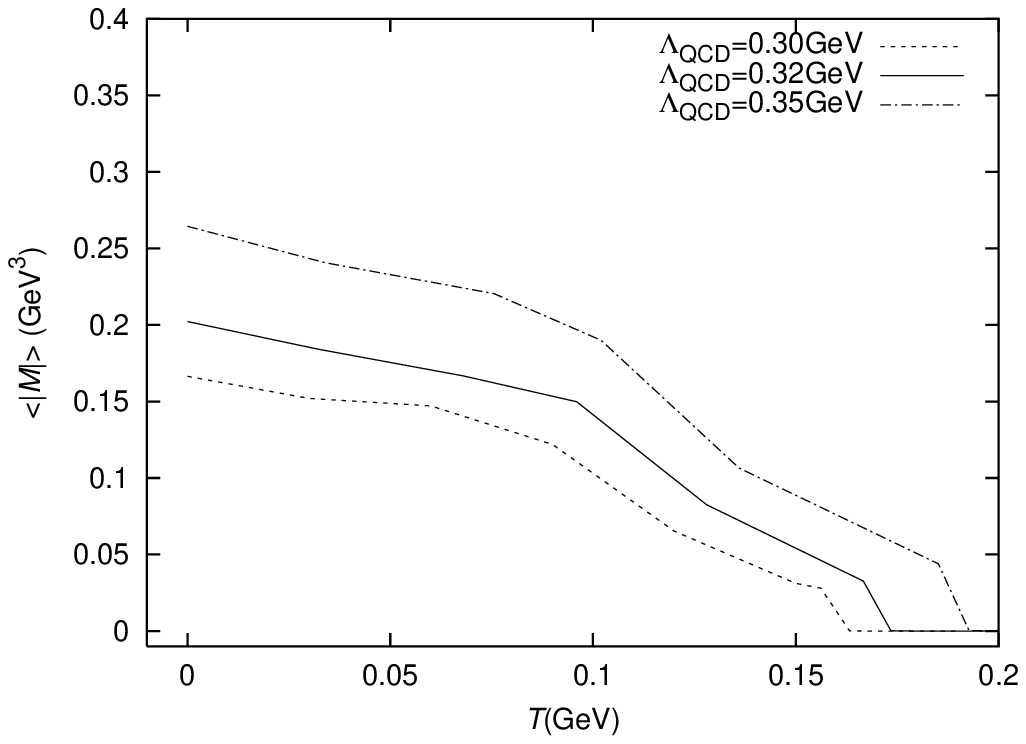}}
\caption{The $T$ dependences of the integrated quark mass functions $\langle |M|\rangle $ with $\Lambda_{\rm QCD}=0.30{\rm GeV}$,$0.32{\rm GeV}$ and $0.35{\rm GeV}$, respectively. }
\end{figure}

\begin{figure}
\centerline{\includegraphics[width=10cm]{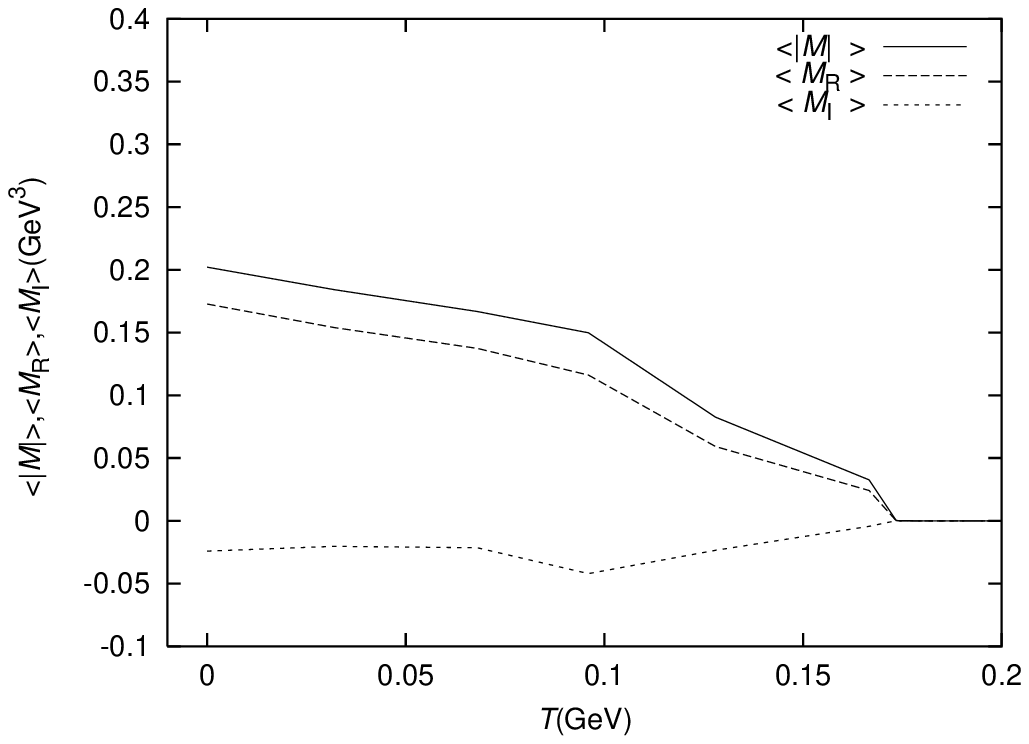}}
\caption{The $T$ dependences of the integrated quark mass functions  $\langle |M|\rangle $,$\langle M_{\rm R}\rangle $ and $\langle M_{\rm I}\rangle $ with $\Lambda_{\rm QCD}=0.32{\rm GeV}$. }
\end{figure}

\section{Summary and comments}

In this paper, we studied quark mass functions solved by the Schwinger-Dyson equation (SDE) in the real-time formalism (RTF) without the instantaneous exchange  
 approximation at finite temperature. The RTF enable us to evaluate non-equilibrium systems. 

 We defined integrated mass functions as order parameters and examined chiral symmetry restoration at finite temperature in the RTF.

In our model, the critical temperature $T_{\rm C}$, in which the chiral symmetry is restored, depends on the QCD scale parameter $\Lambda_{\rm QCD}$. 
Our model roughly gives the critical temperature for the chiral symmetry restoration as $T_{\rm C}/\Lambda_{\rm QCD}\sim 1/2$ with the quark mass for $P^2=(M^2)_R$ at $T=0$ as $\sqrt{(M^2)_R}/\Lambda_{\rm QCD}\sim 1$.
 For example, $\Lambda_{\rm QCD}\simeq 0.32{\rm GeV}$ gives $T_{\rm C}\simeq 0.17{\rm GeV}$ with the effective quark mass $\sqrt{(M^2)_R}\simeq 0.32{\rm GeV}$ for $P^2=(M^2)_R$ at $T=0$. 
 
We  found that the imaginary part of the mass function $\langle M_{\rm I}\rangle $ is non-zero value for broken chiral symmetric phase below  $T_{\rm C}$, which means the massive quark state may be unstable if the energy scale of the system rapidly changes.

@Though our model presented in this paper is too simple to evaluate the quark mass function quantitatively, our study suggests that the SDE in the RTF without the IEA seems to be useful to investigate the chiral phase transition in QCD at finite temperature, in which we can directly evaluate the instability of the massive quark state.

Evaluations of the quark mass function at finite density are difficult task in the imaginary-time formalism   due to sigh problem, particularly Lattice simulations.

On the other hand, in the RTF, the chemical potential may be naturally introduced 
by sifting the energy component.

In future works, we shall extend our method to evaluate the quark mass function at finite density in the RTF.

\section*{Acknowledgements}

This work was supported in part by RCMAS (the Research Center for Measurement in Advanced Science) of Rikkyo University.


\vspace{5mm}


\vspace{5mm}

\begin{center}
{\Large Appendix A :  Formulas for Numerical Calculations}
\end{center}
\vspace{5mm}

The quark mass function is given by 
$$
M^{11}(p_0,p) = -{3iC_{\rm F} \over 2\pi^2}\int^{\Lambda_0}_{-\Lambda_0}dq_0 \int^{\Lambda}_{\delta} dq {q \over p}[\alpha_sMI_{11}J_{11}](p_0,p.q_0,q), 
$$
where  the propagators $ I_{11}$ and $J_{11}$ are defined in Eq.(2$\cdot$9) and Eq.(2$\cdot$10), respectively. \footnote{In order to avoid the infrared contributions due to $M_{\rm I}=(M^{11})_{\rm I}/N_{\rm F}$ and $N_{\rm B}$ for small energy, we regularize the singularity of $\epsilon(x)\coth(x)$ as 
$$ \epsilon(x)\coth(x) \rightarrow \epsilon(x)\coth(x)-\left({1 \over |x|}-{1 \over x_0}\right)\theta(x_0-|x|)$$ with $x_0=0.1$, where $x$ denotes $k_0/(2T)$ or $q_0/(2T)$.
}

As shown in Eq.(2$\cdot$14), we approximate the integration over $q_0$ as
$$
M(p_0,p) \simeq -{3iC_{\rm F} \over 2\pi^2} \int^{\Lambda}_{\delta} dq {q \over p}\sum_{l=1}^{N-1}\langle \alpha_s J_{11}(q) \rangle _l [MI_{11}(q_0^{(l+1)},q_0^{(l)})]
$$
with 
$$
[MI_{11}(q_0^{(l+1)},q_0^{(l)})] = [(MI_{\rm F})(q_0^{(l+1)},q_0^{(l)})]_R + i[(MI_{\rm F})(q_0^{(l+1)},q_0^{(l)})]_IN_{\rm F}(T,\langle q_0\rangle_l),
$$
where, $\langle X\rangle _l$ denotes an average of  $X(q_0^{(l+1)})$ and $X(q_0^{(l)})$ as $ \langle X\rangle _l=[X(q_0^{(l+1)})+X(q_0^{(l)})]/2$.

Here, $[(MI_{\rm F})(q_0^{(l+1)},q_0^{(l)})]_{\rm R} $ and $[(MI_{\rm F})(q_0^{(l+1)},q_0^{(l)})]_{\rm I} $ are given as
$$ [(MI_{\rm F})(q_0^{(l+1)},q_0^{(l)})]_{\rm R} \simeq \langle M_{\rm R}\rangle _l(I_{\rm F}(q_0^{(l+1)},q_0^{(l)}))_{\rm R}-\langle M_{\rm I}\rangle _l(I_{\rm F}(q_0^{(l+1)},q_0^{(l)}))_{\rm I}
$$ 
and
$$ [(MI_{\rm F})(q_0^{(l+1)},q_0^{(l)})]_{\rm I} \simeq \langle M_{\rm R}\rangle _l(I_{\rm F}(q_0^{(l+1)},q_0^{(l)}))_{\rm I}+\langle M_I\rangle _l(I_{\rm F}(q_0^{(l+1)},q_0^{(l)}))_{\rm R}
$$ 
respectively, with
$$
I_{\rm F}(q_0^{(l+1)},q_0^{(l)})=\int^{q_0^{(l+1)}}_{q_0^{(l)}}dq_0I_{\rm F}=(I_{\rm F}(q_0^{(l+1)},q_0^{(l)}))_{\rm R}+i(I_{\rm F}(q_0^{(l+1)},q_0^{(l)}))_{\rm I}.
$$
Here, the strong coupling constant $\alpha_s$ is replaced by the running coupling constant $\alpha_s({\bar P}^2,{\bar Q}^2,T^2)= g^2_s({\bar P}^2,{\bar Q}^2,T^2)/( 4\pi) $[14], which is defined as
$$
g^2_s({\bar P}^2,{\bar Q}^2,T^2)={1 \over \beta_0} \times \left\{ \begin{array}{ll}
     {1 \over t}  & {\rm if}  ~~t_{\rm F} < t \\
     {1 \over t_{\rm F}}+{(t_{\rm F}-t_{\rm C})^2-(t_{\rm C}-t)^2 \over 2t_{\rm F}^ 2(t_{\rm F}-t_{\rm C})}  & {\rm if} ~~ t_{\rm C} < t < t_{\rm F}    \\
      {1 \over t_{\rm F}}+{(t_{\rm F}-t_{\rm C}) \over 2t_{\rm F}^2}  &  {\rm if} ~~t < t_{\rm C}  \end{array} \right\}
$$
with $\beta_0=(33-2N_{\rm f})/(48\pi^2)$,$t=\log[({\bar P}^2+{\bar Q}^2+T^2)/\Lambda^2_{\rm QCD}],t_{\rm F}=0.5$, and $t_{\rm C}=-2$ for $N_{\rm f}=2$ flavors, where, ${\bar P}^2=p_0^2+p^2$ and ${\bar Q}^2=\langle q_0 \rangle_l^2+q^2$.
\footnote{
We implement the QCD coupling constant $\alpha_s$ with Euclidean momenta. Difference between the argument of $\alpha_s$ with the momenta in Minkowski space and that in Euclidean space is a part of higher order contributions to the one-loop approximation. }

For $I_{\rm F}$ and $J_{\rm F}$,  we separate the real parts $(I_{\rm F})_{\rm R},(J_{\rm F})_{\rm R} $ and the imaginary parts $(I_{\rm F})_{\rm I},(J_{\rm F})_{\rm I}$, respectively.

For $J_F$ in Eq.(2$\cdot$12), we can integrate over $k$ as
$$
 (J_F)_{\rm R} = -\int^{\eta_+}_{\eta_-}dk{k(k^2-k_0^2) \over (k^2-k_0^2)^2+\varepsilon^2} =  -{1 \over 4}\log{(\eta_+^2-k_0^2)^2+\varepsilon^2 \over (\eta_-^2-k_0^2)^2+\varepsilon^2} 
$$
 and
$$
 (J_F)_{\rm I}= -\int^{\eta_+}_{\eta_-}dk{k\varepsilon \over  (k^2-k_0^2)^2+\varepsilon^2}    
= -{1 \over 2}\left[\arctan{\eta_+^2-k_0^2 \over \varepsilon}-\arctan{\eta_-^2-k_0^2 \over \varepsilon}\right],
$$ 
respectively, with $\eta_{\pm}=|p\pm q|$ and $k=|{\bf k}|$.

The real  and imaginary parts of the quark propagator $ I_{\rm F}(q_0^{(l+1)},q_0^{(l)})$ are given by         
$$ (I_{\rm F}(q_0^{(l+1)},q_0^{(l)}))_R ={\epsilon(\langle 2q_0-{\partial (E^2)_{\rm R}\over\partial q_0}\rangle _l) \over 2\langle |2q_0-{\partial (E^2)_{\rm R} \over\partial q_0}|\rangle _l } \log{[t_{\rm R}(q_0^{(l+1)},q)]^2+\langle (E^2)_{\rm I}\rangle _l^2 \over [t_{\rm R}(q_0^{(l)}.q)]^2+\langle (E^2)_{\rm I}\rangle _l^2} 
$$
and
$$ (I_{\rm F}(q_0^{(l+1)},q_0^{(l)}))_R={\epsilon(\langle 2q_0-{\partial (E^2)_{\rm R}\over\partial q_0}\rangle _l)\epsilon(\langle (E^2)_{\rm I}\rangle _l) \over \langle |2q_0-{\partial (E^2)_{\rm R} \over\partial q_0}|\rangle _l } $$
$$ \times\left[\arctan{ t_{\rm R}(q_0^{(l+1)},q) \over |\langle (E^2)_{\rm I}\rangle _l|}-\arctan{ t_{\rm R}(q_0^{(l)},q) \over |\langle (E^2)_{\rm I}\rangle _l|}\right], 
$$
respectively, with
$$ t_{\rm R}(q_0,q)=q_0^2-(E^2( q_0,q))_{\rm R} $$
 where, we define $\epsilon(z)=\theta(z)-\theta(-z)$ with the step function $\theta(z)$. Here, the real and imaginary parts of the squared energy denoted by $(E^2)_{\rm R}$ and $(E^2)_{\rm I}$, respectively, are given as
$$
 (E^2)_{\rm R}=q^2+(M^2)_{\rm R} 
$$
and
$$
 (E^2)_{\rm I}=(M^2)_{\rm I}-\varepsilon,
$$
with 
$$
 (M^2)_{\rm R}={\rm Re}(M^2)=(M_{\rm R})^2-(M_{\rm I})^2 
$$
and
$$
 (M^2)_{\rm I}={\rm Im}(M^2)=2M _{\rm R}M_{\rm I}.
$$

\vspace{5mm}


\end{document}